\documentclass[aps,prd,tighten,11pt,showpacs,a4,nofootinbib]{revtex4}

\usepackage{graphicx}
\usepackage{bm}
\usepackage{amsmath,amssymb}
\usepackage{latexsym}
\usepackage{color}
\usepackage{tabularx}

\newcommand{\rp}{r_{+}}

\newcommand{\bee}{\begin{equation}}
\newcommand{\eee}{\end{equation}}

\begin{document}

\title{Vacuum polarization of massive fields in the spacetime of 
the higher-dimensional  black holes.}

\author{Jerzy Matyjasek\email{jurek@kft.umcs.lublin.pl} and Dariusz Tryniecki}

\affiliation{Institute of Physics,
Maria Curie-Sk\l odowska University\\
pl. Marii Curie-Sk\l odowskiej 1,
20-031 Lublin, Poland}

\begin{abstract}
We construct and study the vacuum polarization, $\langle \phi^{2}\rangle_{D},$
of the quantized massive scalar field with a general curvature coupling parameter
in higher-dimensional static and  spherically-symmetric black hole spacetimes, 
with a special 
emphasis put on the electrically charged Tangherlini solutions  and the 
extremal 
and ultraextremal configurations. For $4 \leq D \leq 7$ the explicit analytic 
expressions for the vacuum polarization are given.
For the conformally coupled fields the relation between the trace of the stress-energy 
tensor and the vacuum polarization is examined, which requires knowledge of the
higher-order terms in the Schwinger-DeWitt expansion.
\end{abstract}
\maketitle

\section{Introduction}
The vacuum polarization, $\langle \phi^{2} \rangle,$ has often been regarded as a 
lesser cousin to the stress-energy tensor. However, despite the natural 
interest, there are a few additional reasons to calculate and  study 
$\langle \phi^{2} \rangle .$ Indeed,  it plays an important role in symmetry 
breaking problem and in the calculation of the trace of the stress-energy tensor 
of the conformally coupled quantum fields, being  much easier to find. For 
example, to construct the stress-energy tensor
within the framework of the Schwinger-DeWitt approach, 
one has to functionally differentiate the effective action with respect 
to the metric tensor whereas to calculate the vacuum polarization it is 
sufficient to use the Green function. Moreover, the calculations of the 
vacuum polarization may reveal conceptual subtleties of the problem and help to 
choose the best calculational strategy. Generally speaking, if there are some
technical problems in the calculations of $\langle \phi^{2} \rangle$ the same 
is expected in construction of the stress-energy tensor. On the other hand,
if the calculation of the vacuum polarization goes smoothly the same is expected
for the stress-energy-tensor. In the latter case the only difference is the scale
of the calculations.

Starting with the seminal work of Candelas~\cite{Candelas}, the vacuum
polarization has been, and currently is, studied in a number of physically
interesting 
cases, as for example, Schwarzschild black holes or FRWL cosmologies ~\cite{Candelas,CandelasH,Anderson:1990jh,Frolov:1982pi,Anderson:1989vg,
Fawcett1,
Fawcett2,Kofman,Samuel,Shiraishi0,Quinta,Popov0,Flachi,Tomimatsu,Koyama,
kocio_vac_RN,Popov,Levi}. 
The vacuum polarization effects have been analyzed in the spacetimes
of distorted black holes~\cite{frolov_garcia,frolov_sanchez}, in the spacetimes 
of
dimension higher (or smaller) than 
4~\cite{FrolovPaz,Shiraishi1,Shiraishi2,ThompsonL,jaPhiSq,
MatyjasekFRWL,Quinta0,Taylor1,Taylor2,Flachi2,Breen}. Approximate expressions 
describing $\langle \phi^{2} \rangle$ have been constructed in 
Refs.~\cite{Page,Frolov:1987gw,FZ,Frolovhab,Matyjasek_masslessIHH}.

The aim of this paper is to construct the vacuum polarization of a quantized  massive scalar field 
(with arbitrary curvature coupling) satisfying the covariant
equation
\begin{equation}
 \left(-\Box + m^{2} + \xi R \right) \phi = 0,
\end{equation}
where $m$ is the mass of the field, $\xi$ is the coupling constant
and $R$ is the Ricci scalar, in a general $D$-dimensional static and spherically-symmetric
spacetime described by the line element
\begin{equation}
 ds^{2} = -f(r) dt^{2} + h(r) dr^{2} + r^2 d\Omega^{2}_{D-2},
 \label{ta}
\end{equation}
where $d\Omega^{2}_{D-2}$ is the metric on a unit  
$(D-2)$-dimensional sphere, and to apply the general formulas in the spacetime of the 
charged black hole.
The line element describing such configurations has been constructed by 
Tangherlini in the early sixties.  It has a particularly simple and transparent
form when  parametrized by the radial coordinates of the event and inner horizons, denoted
respectively by $r_{+}$ and $r_{-}.$  The charged Tangherlini solution has the 
form (\ref{ta})
with
\begin{equation}
f(r) = \frac{1}{h(r)} =\left[1-\left( \frac{r_{+}}{r}\right)^{D-3} \right] 
\left[1-\left( 
\frac{r_{-}}{r}\right)^{D-3} \right].
\label{electr}
\end{equation}
Making use of the relations
\begin{equation}
 M = \frac{D-2}{16\pi}\Omega_{D-2}\left(r_{+}^{D-2}+r_{-}^{D-2} \right)
\end{equation}
and
\begin{equation}
 Q =\pm \left(r_{+}r_{-} \right)^{\frac{D-3}{2}}\sqrt{\frac{(D-3)(D-2)}{8\pi}},
\end{equation}
where $\Omega_{D-2}$ is the area of a unit $(D-2)$-sphere, the line element
can be expressed in a standard mass, charge parametrization. The area of the 
$(D-2)$-dimensional sphere is given by
\begin{equation}
 \Omega_{D-2} = \frac{2\pi^{\frac{D-1}{2}}}{\Gamma(\frac{D-1}{2})}.
\end{equation}

When the horizons merge, i. e.,  $(r_{+} = r_{-} = r_{\pm})$  the topology of 
the closest vicinity of the horizon of the extremal black hole is 
$AdS_{2}\times 
S^{D-2},$ and the geometry, when expanded into the whole manifold, is a special 
case of the general solution
\begin{equation}
 ds^{2} = \frac{1}{A}\left(-\sinh^{2}\chi d\tau^{2} + d\chi^{2}\right) + B 
d\Omega^{2}_{D-2},
\label{br1}
\end{equation}
where $A, B \in \mathbb{R}.$ 
It is a product spacetime with maximally symmetric subspaces.
Sometimes it is advantageous
to work with the 
\begin{equation}
 ds^{2} = \frac{1}{A y^{2}}\left(- dT^{2} + dy^{2}\right) + 
B d\Omega^{2}_{D-2}.
\label{br2}
\end{equation}
The equation (\ref{br1}) describes the vicinity of the extremal black hole 
(\ref{ta})
provided  $A = f''(r_{\pm})/2$ and
$B=r_{\pm}^{2}.$ Consequently, 
\begin{equation}
 A = \frac{(D-3)^{2}}{r_{\pm}^{2}}
\end{equation}
and in $D=4$ one has the Bertotti-Robinson solution. 

The calculations of the stress-energy tensor and the vacuum polarization of the 
quantized fields in curved spacetimes are extremely hard as they exhibit a 
nonlocal dependence on the spacetime metric. Here we consider the 
case when the Compton length associated with the field, $\lambda_{c},$ 
satisfies the condition
\begin{equation}
 \frac{\lambda_{c}}{L} \ll 1,
 \label{warunek}
\end{equation}
where $L$ is a characteristic radius of the curvature of the background geometry,
and, consequently, the nonlocal contribution to the vacuum polarization 
can be neglected~\cite{FZ,Matyjasek1,Matyjasek2,jaPhiSq}.  

In the proper-time formalism one assumes that the Green function, $G(x,x'),$ 
that satisfies the equation
\begin{equation}
 \left(\Box - m^{2} - \xi R \right) G(x,x') = -\delta(x,x'),
 \label{eqKG2}
\end{equation}
is given by  
\begin{equation}
G^{F}(x,x') =  \frac{i \Delta^{1/2}}{(4 \pi)^{D/2}} \int_{0}^{\infty} i ds 
\frac{1}{(is)^{D/2}} \exp\left[-i m^{2} s + \frac{i \sigma(x,x')}{2 s} \right] 
A(x,x'; is), 
\label{grf}
\end{equation}
where
\begin{equation}
 A(x,x'; is) = \sum_{k=0}^{\infty} (is)^k a_{k}(x,x'),
\end{equation}
$s$ is the proper time and the biscalars $a_{k}(x,x')$ are the celebrated Hadamard-DeWitt
coefficients, $\Delta(x,x')$ is the vanVleck-Morette determinant and the biscalar $\sigma(x,x')$
is defined as the one-half of the geodetic distance between $x$ and $x'.$
Now, let us define 
\begin{equation}
 A^{(n)}_{reg}(x,x'; is) = A(x,x'; is) - \sum_{k=0}^{[\frac{D}{2}]-1} 
a_{k}(x,x') (is)^k,
\end{equation}
where $\lfloor x \rfloor$ (a floor function)  gives the largest integer 
less than or equal $x,$   substitute  in Eq. (\ref{grf})
$A^{(D)}_{reg}(x,x';is)$ for  $A(x,x'; is)$  and finally denote the thus 
obtained 
biscalar 
by $G^{(D)}_{reg}.$ The field fluctuation that characterizes the vacuum 
polarization in the $D$-dimensional spacetime is defined as
\begin{equation}
 \langle \phi^{2} \rangle = - i \lim_{x' \to x}  G^{(D)}_{reg}.
\end{equation}
Making the substitution
$m^{2} \to m^{2} - i \varepsilon$ ($\varepsilon >0$) the integral (\ref{grf})  can easily 
be calculated~\cite{MatyjasekFRWL,jaPhiSq}:
\begin{equation}
 \langle \phi^{2} \rangle = \frac{1}{(4 \pi)^{D/2}} \sum_{k=\lfloor D/2 
\rfloor}^{N} \frac{a_{k}}{(m^{2})^{k+1-D/2}}  \Gamma\big(k+1-\frac{D}{2}\big),
\label{main}
\end{equation}
where the coincidence limit of the Hadamard-DeWitt biscalars is defined as
$a_{k} = \lim_{x' \to x} a_{k}(x,x')$  
and the upper sum limit, $N,$ remains us that only a first few Hadamard-DeWitt
coefficients are known. One expects that if the condition (\ref{warunek}) holds
Eq.~(\ref{main}) gives a reasonable approximation to the exact $\langle \phi^{2} 
\rangle.$
Eq.~(\ref{main}) is generalization to arbitrary dimension of the formula 
derived by Frolov~\cite{Frolovhab} and coincides with the result obtained in 
Ref.~\cite{ThompsonL}.

The plan of the paper is as follows. In the next section (subsections \ref{sec:cz2a} and 
\ref{sec:cz2b})
we shall construct the vacuum polarization of the quantized massive field 
in the general static and  spherically-symmetric $D$-dimensional spacetime
$(4\leq D \leq 7)$ and use the obtained formulas in the spacetime of the 
charged Tangherlini black holes\footnote{Actually, we have calculated
the vacuum polarization in $4\leq D \leq 9.$ However, the complexity of
the formulas describing $\langle \phi^{2}\rangle_{D}$ rapidly grows 
with $D$ and the results in the higher-dimensional case are rather complicated.
All results can be obtained on request from the first author.}. 
The special emphasis is put on the extremal
and ultraextremal configurations. In the section \ref{sec:cz2c} we shall analyze
the trace of the stress-energy tensor of the conformally coupled massive fields
and analyze its relation to the vacuum polarization. 
The last section~\ref{fin}  concludes the paper with some final remarks, putting our
results in a somewhat broader perspective. Our general results for $\langle 
\phi^{2} \rangle$ in $D$-dimensional  black hole spacetime are relegated to 
Appendix.

Throughout the paper the natural system of unit is used. The signature of the 
metric is `` mainly positive'' $(-,+,...,+)$ and our conventions for curvature 
are $R^{a}_{\ bcd} = \partial_{c} \Gamma^{a}_{bd}  ...$ and $R^{a}_{\ bac} = 
R_{bc}.$

\section{$\langle \phi^{2}\rangle$ in the spacetime of $D$-dimensional
static and spherically-symmetric black hole.}
\label{sec:cz2}
The formula (\ref{main}) shows that the Hadamard-DeWitt coefficients can be 
used in a twofold way: Firstly, for a given dimension the lowest coefficient
of the expansion 
gives the leading approximation to the vacuum polarization, whereas the higher
order coefficients give the higher-order terms in (\ref{main}). On the other hand,
we can confine ourselves only to the main approximation and use the coefficients
in various dimensions. Moreover, for the conformally coupled fields with  $\xi 
= (D-2)/(4 D-4)$ one has a very interesting formula that relates the trace of 
the quantized stress-energy tensor and the vacuum polarization.

In this paper we shall restrict our analyses to 
$4\leq D \leq 7$ and use the first three nontrivial coefficients ($a_{2},
$ $a_{3},$ and $a_{4}$) to calculate all the terms from Table~\ref{ta1}.
Since the results for the higher-order terms as well as these for $D>7$
are rather complicated, to prevent unnecessary proliferation of lengthy and 
not very illuminating formulas,  they will be not presented here.
(The only exception is Sec.~\ref{sec:cz2c}).

As can be seen in Table~\ref{ta1} the possible applications 
of the Hadamard-DeWitt coefficients are of course wider. 
\begin{table}[h!]
\centering
\begin{tabular}{| c | c c c}
\hline
$D$ & 1st & 2nd & 3rd  \\
\hline
4 & $a_{2}$ & $a_{3}$ & $a_{4}$ \\
5 & $a_{2}$ & $a_{3}$ & $a_{4}$ \\
6 & $a_{3}$ & $a_{4}$ \\
7 & $a_{3}$ & $a_{4}$ \\
8 & $a_{4}$ \\
9 & $a_{4}$\\
\end{tabular}
\caption{The rows (from left to right) represent the dimension of the spacetime,
the leading terms, 
the next-to-leading and the  next-to-next-to-leading terms of the 
expansion (\ref{main}).}
\label{ta1}
\end{table}
Indeed, the coefficients $a_{i}$ for $(i \geq 3)$ play a crucial role in
calculations of the stress-energy tensor of the quantized massive fields in a large 
mass limit, giving the unique possibility to study the dependence of the quantum 
effects on the dimension of the background spacetime.
In this case the entries in Tab.~\ref{ta1} should be moved one column to the 
left, as the main approximation of the stress-energy tensor in $D=4$ and 5
requires $a_{3},$ whereas $a_{4}$ is needed in the calculation of the main 
approximation in $D=6$ and 7.

If there are $N$ 
scalar fields $\phi_{i},$ each with a different mass $m_{i},$  then all 
formulas remain intact provided the following change is made
\begin{equation}
 \frac{1}{m^{2}} = \sum_{i=1}^{N} \frac{1}{m_{i}^{2}}.
\end{equation}
This also shows that the quantum effects can be made great by taking
large number of the quantized fields.

\subsection{D = 4 and D = 5}
\label{sec:cz2a}
Inspection of the general formula (\ref{main}) shows that to calculate the 
vacuum polarization of the massive scalar field one needs the coincidence limit 
of the  Hadamard-DeWitt 
coefficient $a_{2},$ which is constructed from the  curvature invariants
$R_{abcd}R^{abcd},$ $R_{ab} R^{ab},$ $R^{2}$ and $\Box R.$  
Although it looks quite simple, the resulting expression for 
$\langle \phi^{2}\rangle $ constructed for a general metric (\ref{ta}) is 
complicated. Indeed, taking $a_{2}$ in the form
\begin{equation}
 a_{2} = \frac{1}{180} R_{abcd} R^{abcd} -\frac{1}{180} R_{ab}R^{ab} + 
\frac{1}{6}\left(\frac{1}{5}-\xi \right) R_{;a}^{\phantom{a} a}+ 
\frac{1}{2}\left( \frac{1}{6}-\xi  \right)^{2} R^{2},
\end{equation}
after some algebra, one has 
\begin{equation}
 \langle \phi^{2}\rangle_{D} = \frac{1}{K} \sum_{l=0}^{2} 
\sum_{k=1}^{13} \alpha_{k}^{i} \xi^{i} F_{k}(r),
\label{og1}
\end{equation}
where 
\begin{equation}
 K = \begin{cases}
      16\pi^{2} m^{2} &\text{if $ D =4$}\\
      32 \pi^{2} m    &\text{if $ D =5$}
     \end{cases}
\end{equation}
and the functions $F_{k}$ (the same for both dimensions) as well as
the dimension-dependent coefficients $\alpha_{k}^{i}$ are shown in 
Table~\ref{ta2a}.

Since the general form of the vacuum polarization can easily be inferred form Table~\ref{ta2a}
it will not be presented explicitly here. Instead, we shall discuss its 
behavior in a few
important regimes. 
First, let us consider the simplest case of the four-dimensional black holes.
On general grounds one expects that for the line element
(\ref{electr})
the result falls as $r^{-6}$
and the most interesting region is the vicinity of the event horizon.
For the physical values of the coupling constant, i.e., for $\xi =0$ and $\xi 
=1/6,$
the general expression calculated at the event horizon reduces to 
\begin{equation}
 K\langle \phi^{2} \rangle_{4} = -\frac{f''}{18 \rp^2}+\frac{1}{60} {f''}^2-\frac{f'}{3 \rp^3}+\frac{13
   {f'}^2}{45 \rp^2}-\frac{1}{30} f^{(3)} f'-\frac{f' f''}{30
   \rp}+\frac{1}{15 \rp^4}
\end{equation}
and
\begin{equation}
  K\langle \phi^{2} \rangle_{4} = \frac{1}{360} {f''}^2+\frac{{f'}^2}{90 \rp^2}-\frac{1}{180} f^{(3)}
   f'-\frac{f' f''}{30 \rp}+\frac{1}{90 \rp^4},
\end{equation}
respectively.

Now, let us assume that the black hole is extremal, i.e., the event and the 
inner horizons coincide and analyze the vacuum polarization on the degenerate 
horizon. It means that $f(r_{\pm}) = f'(r_{\pm}) =0$ and from the previous 
analysis we know that the same result can be obtained calculating the vacuum 
polarization in the product spacetime with the maximally symmetric subspaces.
Inspection of Table~\ref{ta2a} gives the following expression for the vacuum polarization
in the spacetime of the extreme black hole
\begin{equation}
 K \langle \phi^{2} \rangle_{4} =-\frac{f''}{18 r_{\pm}^2}+\xi ^2 \left(-\frac{2 f''}{r_{\pm}^2}
 +\frac{1}{2} f''^2+\frac{2}{r_{\pm}}^4\right)+\xi  \left(\frac{2 f''}{3 
r_{\pm}^2}-\frac{1}{6}
   f''^2-\frac{2}{3 r_{\pm}^4}\right)+\frac{1}{60} f''^2+\frac{1}{15 r_{\pm}^4}.
\end{equation}
If additionally the second derivative of the function $f$ at the event horizon 
vanishes one has the Plebanski-Hacyan geometry with
\begin{equation}
K  \langle \phi^{2} \rangle_{4} = \frac{2 \xi ^2}{\rp^4}-\frac{2 \xi }{3 
\rp^4}+\frac{1}{15 \rp^4}.
\end{equation}

Similarly, at the event horizon  of the five-dimensional black hole one has for 
the minimal 
coupling
\begin{equation}
 K\langle \phi^{2} \rangle_{5} = -\frac{f''}{6 \rp^2}+\frac{1}{60} {f''}^2-\frac{4 
f'}{3 \rp^3}+\frac{59
   {f'}^2}{120 \rp^2}-\frac{1}{30} f^{(3)} f'-\frac{f' f''}{20
   \rp}+\frac{1}{2 \rp^4}
\end{equation}
and for the conformal coupling
\begin{equation}
 K\langle \phi^{2} \rangle_{5} = \frac{1}{64}\left(
-\frac{f''(r)}{6 \rp^2}
+\frac{23}{120} f''^2
+\frac{5 f'}{3 \rp^3}
-\frac{f'^2}{30 \rp^2}
-\frac{2}{15} f^{(3)} f'
-\frac{17 f' f''}{10 \rp}
+\frac{1}{2 \rp^4}
\right).
\end{equation}
The vacuum polarization of the extremal black hole is given by
\begin{equation}
K \langle \phi^{2}\rangle_{5} = -\frac{f''}{6 r_{\pm}^2}+\xi ^2 \left(-\frac{6 
f''}{r_{\pm}^2}+\frac{1}{2}
  { f''}^2+\frac{18}{r_{\pm}^4}\right)+\xi  \left(\frac{2 
f''}{r_{\pm}^2}-\frac{1}{6}
   {f''}^2-\frac{6}{r_{\pm}^4}\right)+\frac{1}{60} {f''}^2+\frac{1}{2 
r_{\pm}^4} 
\end{equation}
whereas for the ultraextremal configuration, one has
\begin{equation}
 K \langle \phi^{2} \rangle_{5} = \frac{18 \xi ^2}{r_{\pm}^4}-\frac{6 \xi 
}{r_{\pm}^4}+\frac{1}{2 r_{\pm}^4}.
\end{equation}
Finally, let us return to the charged black holes and introduce new variables, 
$x$ and $\beta,$ defined as
$x =r/\rp$ and $\beta = r_{-}/r_{+}.$ Simple calculation give
\begin{equation}
  K \langle \phi^{2} \rangle_{4} = \frac{1}{15 \rp^{4}}\left[ 
\frac{1}{x^{6}}\left( 1+ 2 \beta 
  +\beta^{2}\right) -\frac{4}{x^{7}}\left( \beta+\beta^{2} \right) + \frac{13}{3 
x^{8}} \beta^{2}\right]
\label{rn4}
\end{equation}
and
\begin{eqnarray}
   K \langle \phi^{2}\rangle_{5} &=& 
   \frac{1}{5 \rp^{4}}\left[\frac{1}{x^{8}}\left( 2 \beta ^4-4 \beta ^2
   +40 \beta ^2 \xi +2  \right) \right. \nonumber \\
  && \left.  + \frac{1}{x^{10}}\left( 2 \beta ^4+2 \beta ^2-60 \beta ^4 \xi -60 \beta ^2 \xi\right)
   + \frac{1}{x^{12}}\left( -\frac{17 \beta ^4}{6}+10 \beta ^4 \xi ^2+\frac{230 \beta ^4 \xi }{3}\right)
   \right].
   \label{rn5}
\end{eqnarray}
The higher order terms constructed from $a_{3}$ and $a_{4}$ 
can be found in Ref.~\cite{kocio_vac_RN} and the results presented in this 
section generalize those of Lemos and Thompson~\cite{ThompsonL}.

Inspection of Eq.~(\ref{rn4}) shows that that the result (that is independent 
of $\xi$)  is always positive on the event horizon and is tends to $0^{+}$ as 
$x \to \infty,$
whereas in $(D=5)$-case the vacuum polarization exhibits the more complicated 
dependence on the coupling parameter. The asymptotic behavior of the latter is 
shown in Fig. I. 
The vacuum polarization at the event horizon is always nonnegative  for the 
minimal
and the conformal coupling. 
\begin{figure} 
\centering
\includegraphics[width=14cm]{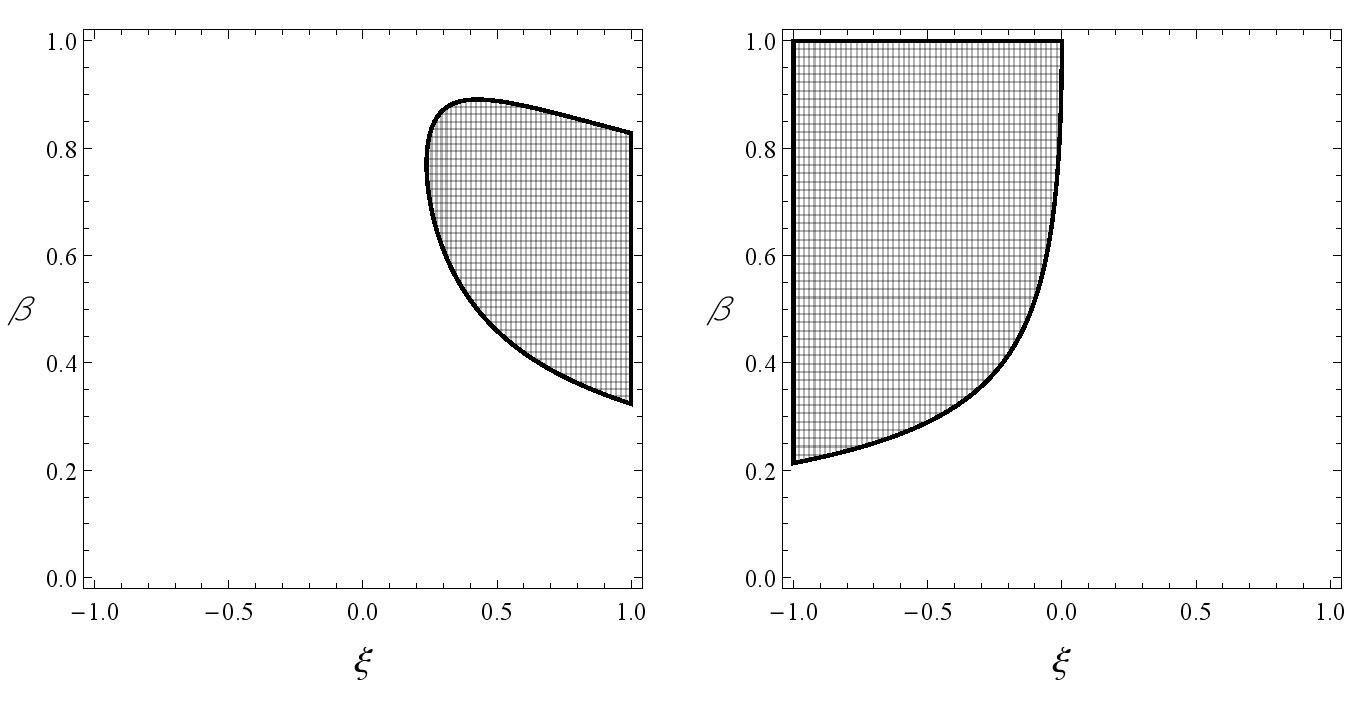}
\caption{ The  shaded regions in the $(\xi, \beta)$-space, in which the field 
fluctuation is negative at the event horizon (left panel). The shaded region in 
the right panel represents points with the property $\langle \phi^{2} 
\rangle_{5} \to 0^{-}$ as $r \to \infty.$  }
\label{rys0}
\end{figure}

Since the second derivative of $f$ calculated at the event horizon is 
$f'' = 2/r_{\pm}^2$ and $f'' = 
8/r_{\pm}^{2}$
one has 
\begin{equation}
 K\langle \phi^{2} \rangle_{4} = \frac{1}{45 r_{\pm}^{2}}
\end{equation}
and
\begin{equation}
 K\langle \phi^{2} \rangle_{5} = \frac{1}{r_{\pm}^{4}} \left( \frac{7}{30} 
-\frac{2}{3}\xi + 2 \xi^{2}\right),
\end{equation}
respectively.

\subsection{D = 6 and D = 7}
\label{sec:cz2b}
In this section we shall analyze the approximation to the field fluctuation of 
the quantized massive fields  in $D=6$ and $D=7.$ The coincidence limit of the 
Hadamard-DeWitt biscalar $a_{3}(x,x')$ is much more complicated than the coefficient
$a_{2}$ and can be written in the form that is valid in any dimension
\begin{equation}
 a_{3} = \frac{1}{7!} b_{3}^{(0)} + \frac{1}{360} b_{3}^{(\xi)},
\end{equation}
where
\begin{eqnarray}
b_{3}^{(0)}&=&
\frac{35}{9}R^3
+17 R_{;a}^{\phantom{;\phantom{a}}}R_{\phantom{;\phantom{a}}}^{;a}
-2 
R_{ab;c}^{\phantom{a}\phantom{b}\phantom{;\phantom{c}}}R_{\phantom{a}\phantom{b}
\phantom{;\phantom{c}}}^{ab;c}
-4 
R_{ab;c}^{\phantom{a}\phantom{b}\phantom{;\phantom{c}}}R_{\phantom{a}\phantom{c}
\phantom{;\phantom{b}}}^{ac;b}
+9 
R_{abcd;e}^{\phantom{a}\phantom{b}\phantom{c}\phantom{d}\phantom{;\phantom{e}}}
R_{\phantom{a}\phantom{b}\phantom{c}\phantom{d}\phantom{;\phantom{e}}}^{abcd;e}
+28 RR_{;a\phantom{a}}^{\phantom{;\phantom{a}}a}
+18 R_{;a\phantom{a}b\phantom{b}}^{\phantom{;\phantom{a}}a\phantom{b}b}
\nonumber \\
&&
-8 
R_{ab;c\phantom{c}}^{\phantom{a}\phantom{b}\phantom{;\phantom{c}}c}R_{\phantom{a
}\phantom{b}}^{ab}
-\frac{14}{3} RR_{ab}^{\phantom{a}\phantom{b}}R_{\phantom{a}\phantom{b}}^{ab}
+24 
R_{ab;c\phantom{b}}^{\phantom{a}\phantom{b}\phantom{;\phantom{c}}b}R_{\phantom{a
}\phantom{c}}^{ac}
-\frac{208}{9} 
R_{ab}^{\phantom{a}\phantom{b}}R_{c\phantom{a}}^{\phantom{c}a}R_{\phantom{b}
\phantom{c}}^{bc}
+12R_{\phantom{a}\phantom{b}\phantom{c}\phantom{d};e\phantom{e}}^{abcd\phantom{
;\phantom{e}}e}R_{abcd}^{\phantom{a}\phantom{b}\phantom{c}\phantom{d}}
\nonumber \\
&&
+ \frac{64}{3} 
R_{ab}^{\phantom{a}\phantom{b}}R_{cd}^{\phantom{c}\phantom{d}}R_{\phantom{a}
\phantom{c}\phantom{b}\phantom{d}}^{acbd}
+ 
\frac{16}{3}R_{ab}^{\phantom{a}\phantom{b}}R_{cde\phantom{a}}^{\phantom{c}
\phantom{d}\phantom{e}a}R_{\phantom{b}\phantom{e}\phantom{c}\phantom{d}}^{becd}
+ \frac{80}{9} 
R_{abcd}^{\phantom{a}\phantom{b}\phantom{c}\phantom{d}}R_{e\phantom{a}f\phantom{
c}}^{\phantom{e}a\phantom{f}c}R_{\phantom{b}\phantom{e}\phantom{d}\phantom{f}}^{
bedf}
+\frac{44}{9} 
R_{abcd}^{\phantom{a}\phantom{b}\phantom{c}\phantom{d}}R_{ef\phantom{a}\phantom{
b}}^{\phantom{e}\phantom{f}ab}R_{\phantom{c}\phantom{d}\phantom{e}\phantom{f}}^{
cdef}\nonumber \\
&&
+ \frac{14}{3} 
RR_{abcd}^{\phantom{a}\phantom{b}\phantom{c}\phantom{d}}R_{\phantom{a}\phantom{b
}\phantom{c}\phantom{d}}^{abcd}
\end{eqnarray}
and
\begin{eqnarray}
b_{3}^{(\xi)}&=&
 -5R^3\xi
 +30R^3\xi^2
 -60R^3\xi^3
 -12\xi R_{;a}^{\phantom{;\phantom{a}}} R_{\phantom{;\phantom{a}}}^{;a}
 +30\xi^2 R_{;a}^{\phantom{;\phantom{a}}}R_{\phantom{;\phantom{a}}}^{;a}
-22R\xi R_{;a\phantom{a}}^{\phantom{;\phantom{a}}a}
+60R\xi^2 R_{;a\phantom{a}}^{\phantom{;\phantom{a}}a}
\nonumber \\
&&
-6\xi R_{;a\phantom{a}b\phantom{b}}^{\phantom{;\phantom{a}}a\phantom{b}b}
-4\xi R_{;ab}^{\phantom{;\phantom{a}}\phantom{b}}R_{\phantom{a}\phantom{b}}^{ab}
+2R\xi R_{ab}^{\phantom{a}\phantom{b}}R_{\phantom{a}\phantom{b}}^{ab}
-2R\xi R_{abcd}^{\phantom{a}\phantom{b}\phantom{c}
\phantom{d}}R_{\phantom{a}\phantom{b}\phantom{c}\phantom{d}}^{abcd}.
\end{eqnarray}
A closer inspection of the coefficient $a_{3}$ shows that it is a sum of the 
curvature invariants constructed from the Riemann tensor, its
covariant derivatives and contractions. In general, the coefficient $[a_{n}]$
(for  a given spin) is a linear combination
of the Riemann invariants and belongs to $\bigoplus_{q=1}^{n} {\cal 
R}_{2n,q}^{0},$
where ${\cal R}_{s,q}^{r}$ is a vector space of Riemannian polynomials of
rank $r$ (the number of free tensor indices), order $s$ (number of derivatives)
and degree $q$ (number of factors). The type of the field is encoded in the 
coefficients of the linear combination.

Now the vacuum fluctuation has a general form
\begin{equation}
 \langle \phi^{2} \rangle_{D} = \frac{1}{K} \sum_{l=0}^{3} \sum_{k=1}^{36} 
\alpha_{k}^{i} \xi^{i} F_{k}(r),
\label{og2}
\end{equation}
where 
\begin{equation}
 K = \begin{cases}
      64\pi^{3} m^{2} &\text{if $ D =6$}\\
      128 \pi^{3} m    &\text{if $ D =7$}
     \end{cases}
\end{equation}
and the functions $F_{k}$ and the (dimension-dependent) coefficients 
$\alpha_{k}$
are listed in Tables~III-V.
Once again we shall not present the general result for $\langle \phi^{2}\rangle_{D}$
as it can easily be obtained form the tables. Instead we shall confine ourselves to
the physically important limits.

Following the steps form the previous section, for the vacuum polarization at 
the 
event horizon of the minimally coupled field one has
\begin{eqnarray}
K\langle \phi^{2}\rangle_{6} &=&-\frac{29 f''}{90 \rp^4}
+\frac{{f''}^2}{30 \rp^2}
-\frac{1}{630} {f''}^3
-\frac{58 f'}{15 \rp^5}
+\frac{116 f'^2}{45 \rp^4}
-\frac{2 f'^3}{315 \rp^3}
-\frac{1}{140} f^{(4)} f'^2 \nonumber \\
&&
-\frac{f^{(3)} f'}{15 \rp^2}
-\frac{8 f^{(3)} f'^2}{315 \rp}
-\frac{f' f''}{15 \rp^3}
+\frac{131 f'^2 f''}{630 \rp^2}
+\frac{2 f' {f''}^2}{315 \rp}
+\frac{1}{210} f^{(3)} f' f''
+\frac{74}{63 \rp^6},
\end{eqnarray}
whereas the analogous result for the  conformally coupled fields is given by
\begin{eqnarray}
K\langle \phi^{2}\rangle_{6} &=&
\frac{f''}{2250 \rp^4}
-\frac{{f''}^2}{750 \rp^2}
-\frac{2 {f''}^3}{7875}
-\frac{2 f'}{375 \rp^5}
-\frac{16 {f'}^2}{1125 \rp^4}
+\frac{2 {f'}^3}{2625 \rp^3}
-\frac{f^{(4)}{f'}^2}{2100}\nonumber \\
&&
-\frac{4 f^{(3)} {f'}^2}{525 \rp}
+\frac{7 f' f''}{375 \rp^3}
-\frac{113 {f'}^2 f''}{15750 \rp^2}
-\frac{13 f' {f''}^2}{7875 \rp}
+\frac{1}{700} f^{(3)}f' f''
-\frac{74}{7875 \rp^6}.
\end{eqnarray}
Usually, the quantum effects of the massive fields are most pronounced
at the event horizon and its closest vicinity.
For the extremal and ultraetremal configurations one has
\begin{eqnarray}
K\langle \phi^{2} \rangle_{6} &=&
-\frac{29 f''}{90 r_{\pm}^4}
+\frac{f''^2}{30 r_{\pm}^2}
-\frac{1}{630} f''^3+\frac{74}{63 r_{\pm}^6}
+\xi ^3 \left(\frac{72 f''}{r_{\pm}^4}
-\frac{6 f''^2}{r_{\pm}^2}+\frac{1}{6} f''^3
-\frac{288}{r_{\pm}^6}\right)
\nonumber \\
&&
+\xi ^2 \left(-\frac{36 f''}{r_{\pm}^4}+\frac{3
f''^2}{r_{\pm}^2}-\frac{1}{12} f''^3+\frac{144}{r_{\pm}^6}\right)
+\xi  \left(\frac{89 f''}{15 r_{\pm}^4}-\frac{8 f''^2}{15
r_{\pm}^2}+\frac{1}{60} f''^3-\frac{116}{5 r_{\pm}^6}\right)
\end{eqnarray}
and
\begin{equation}
K\langle \phi^{2} \rangle_{6} =  -\frac{288 \xi^3}{r_{\pm}^6}+\frac{144 
\xi^2}{r_{\pm}^6}
 -\frac{116 \xi }{5 r_{\pm}^6}+\frac{74}{63 r_{\pm}^6},
\end{equation}
respectively.

\begin{figure} 
\centering
\includegraphics[width=14cm]{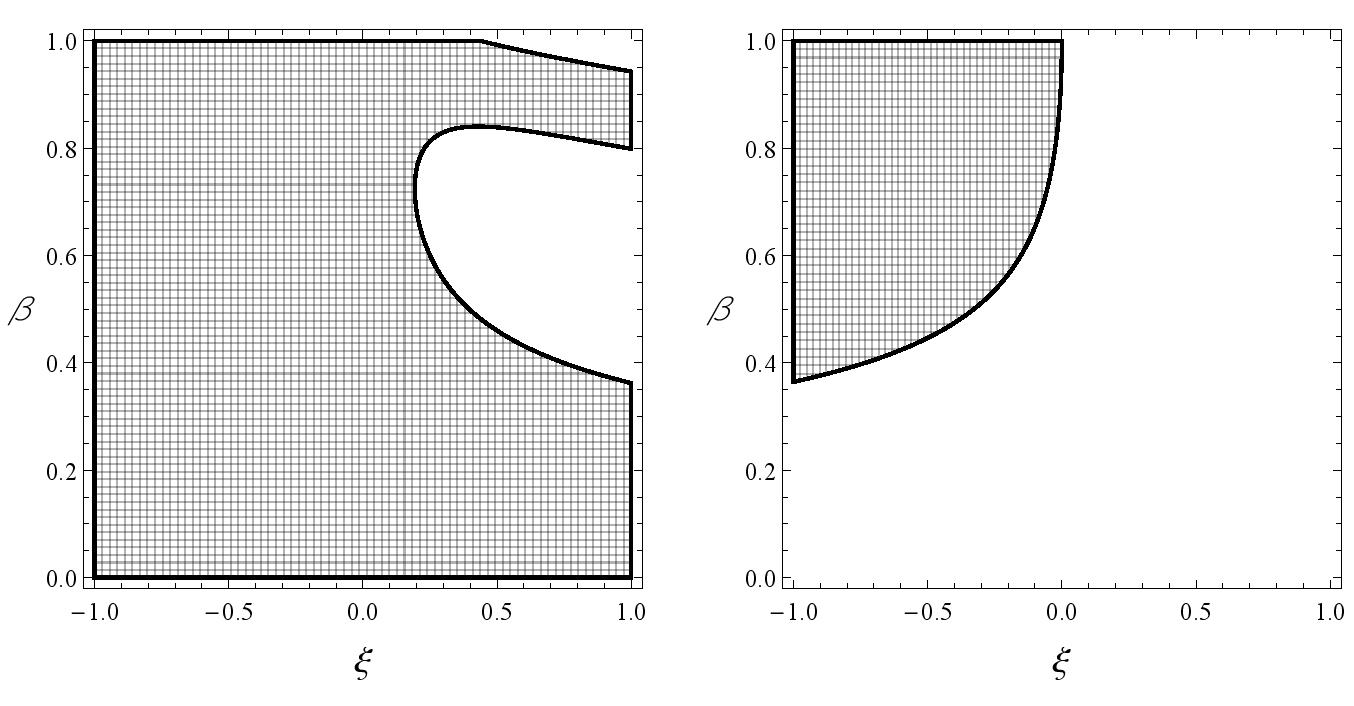}
\caption{ Points within the shaded region represent values of the 
$(\xi,\beta)$ space for which the vacuum polarization is negative at the event
horizon (left panel). The shaded region in 
the right panel represents points with the property $\langle \phi^{2} 
\rangle_{6} \to 0^{-}$ as $r \to \infty.$   }
\label{rys1}
\end{figure}

Thus far our results have been valid for any static and spherically-symmetric metric.
Now let us consider the charged Tangherlini solution. Simple manipulations give
\begin{eqnarray}
\langle \phi^{2} \rangle_{6} &=& \frac{1}{\rp^{6}}\left[
\frac{1}{x^{12}}\left( 15 \beta ^6-30 \beta ^3+280 \beta ^3 \xi +15 \right)
+ \frac{\eta}{x^{15}}\left( -\frac{1333 \beta ^6}{63}+\frac{3382 \beta ^3}{63}
-1232 \beta ^3 \xi -\frac{1333}{63} \right) \right.
\nonumber \\
&& + \frac{\beta^{3}}{x^{18}}\left( \frac{907 \beta ^6}{105}
-\frac{1250 \beta^3}{21}
+432 \beta ^3 \xi^2
+\frac{5448 \beta^6 \xi }{5}
+\frac{21576 \beta ^3 \xi }{5}+\frac{5448 \xi }{5}
+\frac{907}{105} \right)\nonumber \\
&&\left. 
+ \frac{\beta^{6} \eta}{x^{21}}\left( -576 \xi ^2
-\frac{16736 \xi }{5}-\frac{88}{35}\right)
+ \frac{\beta^{9}}{x^{24}} \left( 36 \xi ^3+702 \xi ^2+2392 \xi 
+\frac{6761}{315}\right)\right],
\end{eqnarray}
where $\eta = 1+\beta^{3}.$ The sign of the vacuum polarization at the event horizon
as well as its asymptotic behavior as $r \to \infty$ is shown in Fig.~\ref{rys1}. Specifically,
the vacuum polarization at the event horizon is always negative for the minimal coupling.
On the other hand, for the conformal coupling, it is positive for $0.683< \xi 
<0.762.$

\begin{figure} 
\centering
\includegraphics[width=14cm]{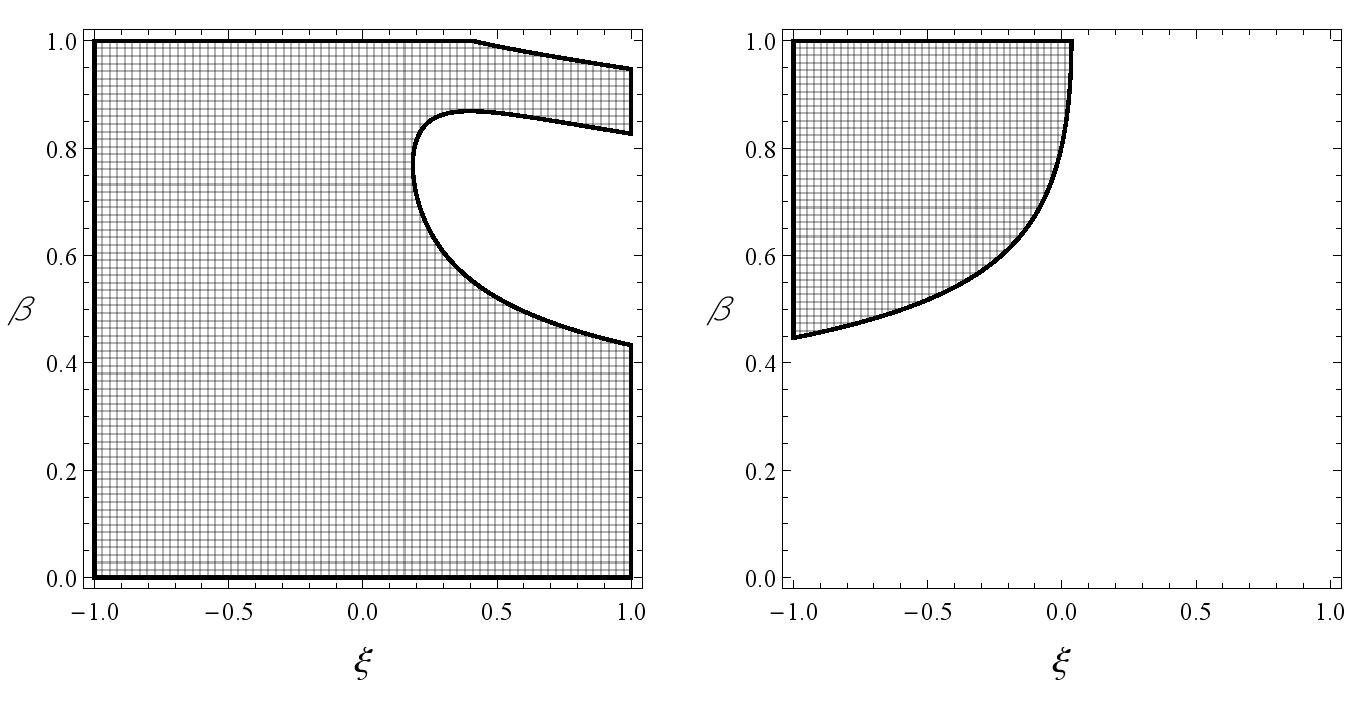}
\caption{ Points within the shaded region represent values of the 
$(\xi,\beta)$ space for which the vacuum polarization is negative at the event
horizon (left panel). The shaded region in 
the right panel represents points with the property $\langle \phi^{2} 
\rangle_{7} \to 0^{-}$ as $r \to \infty.$  }
\label{rys2}
\end{figure}

Now, let us analyze the results obtained for the $7$-dimensional black hole.
At the event horizon for the conformal and minimal coupling  one has
\begin{eqnarray}
K\langle \phi^{2}\rangle_{7} &=&
\frac{f''}{192 \rp^4}
-\frac{7 {f''}^2}{2304 \rp^2}
-\frac{653 {f''}^3}{2903040}
-\frac{f'}{32 \rp^5}
-\frac{13 {f'}^2}{1728 \rp^4}
-\frac{221 {f'}^3}{72576\rp^3}
-\frac{f^{(4)} {f'}^2}{5040}
-\frac{f^{(3)} f'}{864 \rp^2}
\nonumber \\
&&
-\frac{23 f^{(3)} {f'}^2}{4032 \rp}
+\frac{47 f' f''}{1728 \rp^3}
-\frac{139 {f'}^2 f''}{16128 \rp^2}
-\frac{37 f' {f''}^2}{96768 \rp}
+\frac{191 f^{(3)} f' f''}{120960}
-\frac{13}{1296 \rp^6}
\end{eqnarray}
and
\begin{eqnarray}
K\langle \phi^{2}\rangle_{7} &=&
-\frac{8 f''}{9 \rp^4}
+\frac{{f''}^2}{18 \rp^2}
-\frac{1}{630} {f''}^3
-\frac{112 f'}{9 \rp^5}
+\frac{199 {f'}^2}{36 \rp^4}
+\frac{19 {f'}^3}{126 \rp^3}
-\frac{1}{140} f^{(4)} {f'}^2
\nonumber \\
&&
-\frac{f^{(3)} f'}{9 \rp^2}
-\frac{2 f^{(3)} {f'}^2}{63 \rp}
-\frac{f' f''}{6 \rp^3}
+\frac{97 {f'}^2 f''}{336 \rp^2}
+\frac{f' {f''}^2}{126 \rp}
+\frac{1}{210} f^{(3)} f' f''
+\frac{16}{3 \rp^6},
\end{eqnarray}
respectively.
On the other hand, the vacuum polarization for the extremal and ultraextremal
black hole is given by
\begin{eqnarray}
K\langle\phi^{2}\rangle_{7} &=&
-\frac{8 f''}{9 r_{\pm}^4}
+\frac{{f''}^2}{18 r_{\pm}^2}
-\frac{1}{630} {f''}^3
+\frac{16}{3 r_{\pm}^6}
+
\frac{\xi ^3 \left(r_{\pm}^2 f''-20\right)^3}{6 r_{\pm}^6}
-\frac{\xi ^2 \left(r_{\pm}^2 f''-20\right)^3}{12 r_{\pm}^6}\nonumber \\
&&
+\frac{\xi  \left(3 r_{\pm}^4 {f''}^2-100 r_{\pm}^2 f''+960\right)
   \left(r_{\pm}^2 f''-20\right)}{180 r_{\pm}^6}
\end{eqnarray}
and
\begin{equation}
 K \langle \phi^{2} \rangle_{7} = -\frac{16 (5 \xi -1)^2 (10 \xi -1)}{3 r^6}.
\end{equation}
And finally the result for the electrically charged Tangherlini black hole is
\begin{eqnarray}
\langle \phi^{2} \rangle_{7} &=& \frac{1}{\rp^{6}}\left[
\frac{1}{x^{14}}\left(\frac{360 \beta ^8}{7}
-144 \beta ^4+1152 \beta ^4 \xi +\frac{360}{7} \right)
+\frac{24\eta}{x^{18}}\left(\frac{109 \beta ^4}{7}-232 
\beta ^4 \xi 
-\frac{22}{7} -\frac{22 \beta ^8}{7}\right)\right.
\nonumber \\
&&
+ \frac{\beta^{4}}{x^{22}}\left(-\frac{940 \beta ^8}{7}
-\frac{49688 \beta ^4}{63}+2640 \beta ^4 \xi ^2
+5160 \beta ^8 \xi +\frac{60944 \beta ^4 \xi }{3}+5160 \xi
   -\frac{940}{7} \right)
\nonumber \\
&&\left. + \frac{\eta \beta^{8}}{x^{26}}
\left( -3600 \xi ^2 -\frac{48400 \xi }{3}
+\frac{1240}{3}\right) 
+
\frac{\beta^{12}}{x^{30}}\left(288 \xi ^3
+4416 \xi ^2
+\frac{175072 \xi }{15}-\frac{60544}{315}\right) \right],\nonumber\\
\end{eqnarray}
where $\eta =1 + \beta^{4}.$
The sign of the vacuum polarization at the event horizon
as well as its asymptotic behavior as $r \to \infty$ is shown in Fig.~\ref{rys2}.
A more close examination shows that
the vacuum polarization at the event horizon is always negative for the minimal coupling,
whereas for the conformal coupling it is positive for $0.697< \xi <0.825.$ Note 
qualitative similarity of the results presented in Fig~\ref{rys1} to 
Fig.~\ref{rys2}.

Finally, we remark that the vacuum polarization on the degenerate 
horizon  of the extremal black hole can easily be calculated using
the line element (\ref{br1}) describing the spacetimes with the maximally symmetric
subspaces. Because of the symmetries this approach is especially useful 
for the higher-dimensional black holes.

\subsection{Trace of the stress-energy tensor of the conformally coupled 
massive fields}
\label{sec:cz2c}
The one-loop effective action constructed from the Green function (\ref{grf}) 
is given by
\begin{equation}
 W^{(1)} = \int d^{D}x (-g)^{1/2} {\cal L},
\end{equation}
where
\begin{equation}
 {\cal L} = \frac{1}{2 (4 \pi)^{D/2}} \sum_{k= \lfloor \frac{D}{2} \rfloor 
+1}^{N}  \frac{a_{k}}{(m^{2})^{k-D/2}} \Gamma(k-\frac{D}{2}),
\end{equation}
and the stress-energy tensor can be calculated using the standard formula
\begin{equation}
 T^{ab} = \frac{2}{(-g)^{1/2}} \frac{\delta}{\delta g_{ab}} W^{(1)}.
 \label{fun_deriv}
\end{equation}
It should be noted that the total divergences that are present in the
effective action can be discarded. For example, when calculating the 
stress-energy tensor in $D=4$ and $D=5$ the number of the curvature invariants
can be reduced from 28 to 10.

For the conformally coupled fields one has interesting relations between
the trace of the stress-energy tensor and the field fluctuation. Indeed,
provided $\xi = (D-2)/(4 D-4),$ we have
\begin{equation}
 \langle T_{a}^{a} \rangle_{D} = \begin{cases}
       \mathfrak{C}_{D} - m^{2} \langle \phi^{2} \rangle_{D} &\text{for  $ D$-even}\\
      \frac{1}{2}m \mathfrak{C}_{D} - m^{2} \langle \phi^{2} \rangle_{D}  &\text{for $ D$-odd},
     \end{cases}
     \label{trace}
\end{equation}
where $ \mathfrak{C}_{D}$ is given by
\begin{equation}
 \mathfrak{C}_{D} = \frac{a_{\lfloor D/2 \rfloor}}{(4 \pi)^{\lfloor D/2 \rfloor}}.
\end{equation}
For the Schwinger-DeWitt expansion the first-order term cancels with the ``anomalous term''
and the next to leading term is precisely the first-order approximation to the trace.
Similarly, the next-to leading term of the trace is equal to the next-to-next-to leading term of
the vacuum polarization $\langle \phi^{2} \rangle_{D}.$
It should be noted that although the calculation of the trace of the stress-energy with the aid of 
Eq.~(\ref{trace}) requires prior knowledge of the next-to-leading terms of the field fluctuation
(which are expressed in terms of the Hadamard-DeWitt coefficients),
it is still much more simple than the computations of the functional derivatives of the action 
with respect to the metric tensor. Moreover, it can be regarded as a useful check of the
calculations.

Using (\ref{fun_deriv}) we have constructed the stress-energy 
tensor of the massive quantized 
fields in $4 \leq D\leq 7$ for the general static and spherically-symmetric
spacetime.Additionally,   for $D=4$ and 5 we have also calculated 
the next-to-leading terms. On the other hand, we have calculated the first
three terms of the expansion of the field fluctuation in $D=4$ and $D=5$ and 
the first two terms in $D=6$ and $D=7$ and checked the validity of  Eq.~(\ref{trace})
for the general static and spherically-symmetric metric. Since the final results 
are complicated and not very illuminating, here we present only the trace of the stress-energy tensor 
in the spacetime of the Tangherlini black hole. Regardless of the 
adapted method, after some algebra, one has

\begin{equation}
\langle T_{a}^{a} \rangle_{4} =-\frac{\rp^2 (81 r-97 \rp)}{4032 \pi ^2 m^2 
r^9}-\frac{\rp^2 \left(420 r^2-1125 r \rp+727 \rp^2\right)}{2100 \pi ^2 m^4, 
r^{12}}
\end{equation}
\begin{equation}
\langle T_{a}^{a} \rangle_{5} =-\frac{\rp^4 \left(243 r^2-323 \rp^2\right)}{5040 
\pi ^2 m r^{12}}-\frac{\rp^4 \left(1152 r^4-3704 r^2 \rp^2+2727 
\rp^4\right)}{2240 \pi ^2 m^3 r^{16}},
\end{equation}
\begin{equation}
\langle T_{a}^{a} \rangle_{6} =-\frac{\rp^6 \left(22680 r^6-81970 r^3 
\rp^3+65041 \rp^6\right)}{10080 \pi ^3 m^2 r^{20}}
\end{equation}
and
\begin{equation}
\langle T_{a}^{a} \rangle_{7} =-\frac{\rp^8 \left(320 r^8-1256 r^4 \rp^4+1047 
\rp^8\right)}{128 \pi ^3 m r^{24}}.
\end{equation}

\section{Final remarks}
\label{fin}
The Schwinger-DeWitt method gives unique possibility to study 
the quantum effects in various dimensions.
Moreover, as the sole criterion for its applicability is
demanding that the  Compton  length associated with the field be small
the with respect to the characteristic radius 
of the curvature of the background geometry, the Schwinger-DeWitt approach
is also quite robust.  
In practice, it turns out that the reasonable results can be obtained for $ M m > 2$,
where $M$ is the mass of the black hole~\cite{TPaul}. A comparison made 
in Refs.~\cite{BreenO,ThompsonL} between the numerical and analytical results 
confirms accuracy of the Schwinger-DeWitt method. 

In this paper, using the generalized Schwinger-DeWitt approach,
we have calculated the vacuum polarization effects of the 
quantized massive scalar field in the spacetime of the $D$-dimensional 
$(4\leq D \leq 7)$ static and spherically-symmetric black hole. A special 
emphasis has been put on the charged Tangherlini solutions.
Contrary to the simplest Reissner-Nordstr\"om case,
the vacuum polarization for the charged Tangherlini black hole
depends on $\xi$ and a ratio $r_{-}/r_{+}$  in a quite complicated way, as 
expected.
Our results can also be used to construct $\langle \phi^{2} \rangle$ when a 
cosmological constant is present, as, for example, in a spacetime of the 
lukewarm black hole. They can be generalized to the case of topological black 
holes.

Since the geometry of the closest vicinity of the extremal
black hole is a direct product of the two maximally symmetric spaces 
$AdS_{2} \times S^{D-2}$
it is possible to calculate $\langle \phi^{2} \rangle_{D}$ at the 
degenerate horizon without referring to the black hole metric.
Because of massive simplifications in the product space, the thus
constructed result is relatively simple to obtain and may serve as an important check
of the calculations. 

For the conformally coupled field we have investigated the relation 
between the trace of the stress-energy tensor and the vacuum polarization.
It should be emphasized that the  calculation of the trace from the vacuum 
polarization
is far more efficient than the calculations of the trace from the stress-energy tensor,
even though the next-to-leading terms of the approximation of $\langle \phi^{2}\rangle_{D}$
are needed.

Finally, we briefly describe our calculational strategy. First, we
have constructed the Hadamard-DeWitt coefficients for a general $D$-dimensional
metric~(\ref{ta}). The hard part of the calculations has been carried out using FORM~\cite{jos}
(a well-known program in high-energy physics), 
whereas massive simplifications have been performed in MATHEMATICA. Further, 
the functional
derivatives of the general effective action (constructed from the complicated algebraic 
and differential curvature invariants)
with respect to the metric tensor 
tensor have been calculated using fast FORM code. The results have been checked 
against the analogous results obtained from the time and radial Euler-Lagrange equations.
The remaining independent component of the stress-energy tensor has been constructed
with the aid of the covariant conservation equation.

\begin{acknowledgments}
J.M. was partially supported by the Polish National Science Centre grant no. 
DEC-2014/15/B/ST2/00089.
\end{acknowledgments}



\begin{thebibliography}{41}
\expandafter\ifx\csname natexlab\endcsname\relax\def\natexlab#1{#1}\fi
\expandafter\ifx\csname bibnamefont\endcsname\relax
  \def\bibnamefont#1{#1}\fi
\expandafter\ifx\csname bibfnamefont\endcsname\relax
  \def\bibfnamefont#1{#1}\fi
\expandafter\ifx\csname citenamefont\endcsname\relax
  \def\citenamefont#1{#1}\fi
\expandafter\ifx\csname url\endcsname\relax
  \def\url#1{\texttt{#1}}\fi
\expandafter\ifx\csname urlprefix\endcsname\relax\def\urlprefix{URL }\fi
\providecommand{\bibinfo}[2]{#2}
\providecommand{\eprint}[2][]{\url{#2}}
\bibitem[{\citenamefont{Candelas}(1980)}]{Candelas}
\bibinfo{author}{\bibfnamefont{P.}~\bibnamefont{Candelas}},
  \bibinfo{journal}{Phys. Rev.} \textbf{\bibinfo{volume}{D21}},
  \bibinfo{pages}{2185} (\bibinfo{year}{1980}).
\bibitem[{\citenamefont{Candelas and Howard}(1984)}]{CandelasH}
\bibinfo{author}{\bibfnamefont{P.}~\bibnamefont{Candelas}} \bibnamefont{and}
  \bibinfo{author}{\bibfnamefont{K.~W.} \bibnamefont{Howard}},
  \bibinfo{journal}{Phys. Rev.} \textbf{\bibinfo{volume}{D29}},
  \bibinfo{pages}{1618} (\bibinfo{year}{1984}).
\bibitem[{\citenamefont{Anderson}(1990)}]{Anderson:1990jh}
\bibinfo{author}{\bibfnamefont{P.~R.} \bibnamefont{Anderson}},
  \bibinfo{journal}{Phys. Rev.} \textbf{\bibinfo{volume}{D41}},
  \bibinfo{pages}{1152} (\bibinfo{year}{1990}).
\bibitem[{\citenamefont{Frolov}(1982)}]{Frolov:1982pi}
\bibinfo{author}{\bibfnamefont{V.~P.} \bibnamefont{Frolov}},
  \bibinfo{journal}{Phys. Rev.} \textbf{\bibinfo{volume}{D26}},
  \bibinfo{pages}{954} (\bibinfo{year}{1982}).
\bibitem[{\citenamefont{Anderson}(1989)}]{Anderson:1989vg}
\bibinfo{author}{\bibfnamefont{P.~R.} \bibnamefont{Anderson}},
  \bibinfo{journal}{Phys. Rev.} \textbf{\bibinfo{volume}{D39}},
  \bibinfo{pages}{3785} (\bibinfo{year}{1989}).
\bibitem[{\citenamefont{Fawcett and Whiting}(1981)}]{Fawcett1}
\bibinfo{author}{\bibfnamefont{M.~S.} \bibnamefont{Fawcett}} \bibnamefont{and}
  \bibinfo{author}{\bibfnamefont{B.~F.} \bibnamefont{Whiting}}, in
  \emph{\bibinfo{booktitle}{{Nuffield Workshop on Quantum Structure of Space
  and Time }}} (\bibinfo{year}{1981}), pp.
  \bibinfo{pages}{131--154}.
\bibitem[{\citenamefont{Fawcett}(1983)}]{Fawcett2}
\bibinfo{author}{\bibfnamefont{M.~S.} \bibnamefont{Fawcett}},
  \bibinfo{journal}{Commun. Math. Phys.} \textbf{\bibinfo{volume}{89}},
  \bibinfo{pages}{103} (\bibinfo{year}{1983}).
\bibitem[{\citenamefont{Kofman and Sahni}(1983)}]{Kofman}
\bibinfo{author}{\bibfnamefont{L.~A.} \bibnamefont{Kofman}} \bibnamefont{and}
  \bibinfo{author}{\bibfnamefont{V.}~\bibnamefont{Sahni}},
  \bibinfo{journal}{Phys. Lett.} \textbf{\bibinfo{volume}{127B}},
  \bibinfo{pages}{197} (\bibinfo{year}{1983}).
\bibitem[{\citenamefont{Anderson et~al.}(1995)\citenamefont{Anderson, Hiscock,
  and Samuel}}]{Samuel}
\bibinfo{author}{\bibfnamefont{P.~R.} \bibnamefont{Anderson}},
  \bibinfo{author}{\bibfnamefont{W.~A.} \bibnamefont{Hiscock}},
  \bibnamefont{and} \bibinfo{author}{\bibfnamefont{D.~A.}
  \bibnamefont{Samuel}}, \bibinfo{journal}{Phys. Rev.}
  \textbf{\bibinfo{volume}{D51}}, \bibinfo{pages}{4337} (\bibinfo{year}{1995}).
\bibitem[{\citenamefont{Shiraishi}(1992)}]{Shiraishi0}
\bibinfo{author}{\bibfnamefont{K.}~\bibnamefont{Shiraishi}},
  \bibinfo{journal}{Mod. Phys. Lett.} \textbf{\bibinfo{volume}{A7}},
  \bibinfo{pages}{3569} (\bibinfo{year}{1992}).
\bibitem[{\citenamefont{Quinta et~al.}(2016)\citenamefont{Quinta, Flachi, and
  Lemos}}]{Quinta}
\bibinfo{author}{\bibfnamefont{G.~M.} \bibnamefont{Quinta}},
  \bibinfo{author}{\bibfnamefont{A.}~\bibnamefont{Flachi}}, \bibnamefont{and}
  \bibinfo{author}{\bibfnamefont{J.~P.~S.} \bibnamefont{Lemos}},
  \bibinfo{journal}{Phys. Rev.} \textbf{\bibinfo{volume}{D93}},
  \bibinfo{pages}{124073} (\bibinfo{year}{2016}).
\bibitem[{\citenamefont{Popov}(2004)}]{Popov0}
\bibinfo{author}{\bibfnamefont{A.~A.} \bibnamefont{Popov}},
  \bibinfo{journal}{Phys. Rev.} \textbf{\bibinfo{volume}{D70}},
  \bibinfo{pages}{084047} (\bibinfo{year}{2004}.
\bibitem[{\citenamefont{Flachi and Tanaka}(2008)}]{Flachi}
\bibinfo{author}{\bibfnamefont{A.}~\bibnamefont{Flachi}} \bibnamefont{and}
  \bibinfo{author}{\bibfnamefont{T.}~\bibnamefont{Tanaka}},
  \bibinfo{journal}{Phys. Rev.} \textbf{\bibinfo{volume}{D78}},
  \bibinfo{pages}{064011} (\bibinfo{year}{2008}).
\bibitem[{\citenamefont{Tomimatsu and Koyama}(2000)}]{Tomimatsu}
\bibinfo{author}{\bibfnamefont{A.}~\bibnamefont{Tomimatsu}} \bibnamefont{and}
  \bibinfo{author}{\bibfnamefont{H.}~\bibnamefont{Koyama}},
  \bibinfo{journal}{Phys. Rev.} \textbf{\bibinfo{volume}{D61}},
  \bibinfo{pages}{124010} (\bibinfo{year}{2000}).
\bibitem[{\citenamefont{Koyama et~al.}(2000)\citenamefont{Koyama, Nambu, and
  Tomimatsu}}]{Koyama}
\bibinfo{author}{\bibfnamefont{H.}~\bibnamefont{Koyama}},
  \bibinfo{author}{\bibfnamefont{Y.}~\bibnamefont{Nambu}}, \bibnamefont{and}
  \bibinfo{author}{\bibfnamefont{A.}~\bibnamefont{Tomimatsu}},
  \bibinfo{journal}{Mod. Phys. Lett.} \textbf{\bibinfo{volume}{A15}},
  \bibinfo{pages}{815} (\bibinfo{year}{2000}).
\bibitem[{\citenamefont{Matyjasek et~al.}(2010)\citenamefont{Matyjasek,
  Tryniecki, and Zwierzchowska}}]{kocio_vac_RN}
\bibinfo{author}{\bibfnamefont{J.}~\bibnamefont{Matyjasek}},
  \bibinfo{author}{\bibfnamefont{D.}~\bibnamefont{Tryniecki}},
  \bibnamefont{and}
  \bibinfo{author}{\bibfnamefont{K.}~\bibnamefont{Zwierzchowska}},
  \bibinfo{journal}{Phys. Rev.} \textbf{\bibinfo{volume}{D81}},
  \bibinfo{pages}{124047} (\bibinfo{year}{2010}).
\bibitem[{\citenamefont{Popov}(2016)}]{Popov}
\bibinfo{author}{\bibfnamefont{A.~A.} \bibnamefont{Popov}},
  \bibinfo{journal}{Phys. Rev.} \textbf{\bibinfo{volume}{D94}},
  \bibinfo{pages}{124033} (\bibinfo{year}{2016}).
\bibitem[{\citenamefont{Levi and Ori}(2016)}]{Levi}
\bibinfo{author}{\bibfnamefont{A.}~\bibnamefont{Levi}} \bibnamefont{and}
  \bibinfo{author}{\bibfnamefont{A.}~\bibnamefont{Ori}},
  \bibinfo{journal}{Phys. Rev.} \textbf{\bibinfo{volume}{D94}},
  \bibinfo{pages}{044054} (\bibinfo{year}{2016}).
\bibitem[{\citenamefont{Frolov and Garcia}(1983)}]{frolov_garcia}
\bibinfo{author}{\bibfnamefont{V.~P.}~\bibnamefont{Frolov}} \bibnamefont{and}
  \bibinfo{author}{\bibfnamefont{A.}~ \bibnamefont{Garcia}},
  \bibinfo{journal}{Physics Letters A} \textbf{\bibinfo{volume}{99}},
  \bibinfo{pages}{421 } (\bibinfo{year}{1983}).
\bibitem[{\citenamefont{Frolov and Sanchez}(1986)}]{frolov_sanchez}
\bibinfo{author}{\bibfnamefont{V.~P.} \bibnamefont{Frolov}} \bibnamefont{and}
  \bibinfo{author}{\bibfnamefont{N.~G.} \bibnamefont{Sanchez}},
  \bibinfo{journal}{Phys. Rev.} \textbf{\bibinfo{volume}{D33}},
  \bibinfo{pages}{1604} (\bibinfo{year}{1986}).
\bibitem[{\citenamefont{Frolov et~al.}(1989)\citenamefont{Frolov, Mazzitelli,
  and Paz}}]{FrolovPaz}
\bibinfo{author}{\bibfnamefont{V.~P.} \bibnamefont{Frolov}},
  \bibinfo{author}{\bibfnamefont{F.~D.} \bibnamefont{Mazzitelli}},
  \bibnamefont{and} \bibinfo{author}{\bibfnamefont{J.~P.} \bibnamefont{Paz}},
  \bibinfo{journal}{Phys. Rev.} \textbf{\bibinfo{volume}{D40}},
  \bibinfo{pages}{948} (\bibinfo{year}{1989}).
\bibitem[{\citenamefont{Shiraishi and Maki}(1994{\natexlab{a}})}]{Shiraishi1}
\bibinfo{author}{\bibfnamefont{K.}~\bibnamefont{Shiraishi}} \bibnamefont{and}
  \bibinfo{author}{\bibfnamefont{T.}~\bibnamefont{Maki}},
  \bibinfo{journal}{Class. Quant. Grav.} \textbf{\bibinfo{volume}{11}},
  \bibinfo{pages}{695} (\bibinfo{year}{1994}{\natexlab{a}}).
\bibitem[{\citenamefont{Shiraishi and Maki}(1994{\natexlab{b}})}]{Shiraishi2}
\bibinfo{author}{\bibfnamefont{K.}~\bibnamefont{Shiraishi}} \bibnamefont{and}
  \bibinfo{author}{\bibfnamefont{T.}~\bibnamefont{Maki}},
  \bibinfo{journal}{Class. Quant. Grav.} \textbf{\bibinfo{volume}{11}},
  \bibinfo{pages}{1687} (\bibinfo{year}{1994}{\natexlab{b}}).
\bibitem[{\citenamefont{Thompson and Lemos}(2009)}]{ThompsonL}
\bibinfo{author}{\bibfnamefont{R.~T.} \bibnamefont{Thompson}} \bibnamefont{and}
  \bibinfo{author}{\bibfnamefont{J.~P.~S.} \bibnamefont{Lemos}},
  \bibinfo{journal}{Phys. Rev.} \textbf{\bibinfo{volume}{D80}},
  \bibinfo{pages}{064017} (\bibinfo{year}{2009}).
\bibitem[{\citenamefont{Matyjasek and Sadurski}(2015)}]{jaPhiSq}
\bibinfo{author}{\bibfnamefont{J.}~\bibnamefont{Matyjasek}} \bibnamefont{and}
  \bibinfo{author}{\bibfnamefont{P.}~\bibnamefont{Sadurski}},
  \bibinfo{journal}{Phys. Rev.} \textbf{\bibinfo{volume}{D91}},
  \bibinfo{pages}{044027} (\bibinfo{year}{2015}).
\bibitem[{\citenamefont{Matyjasek and Sadurski}(2014)}]{MatyjasekFRWL}
\bibinfo{author}{\bibfnamefont{J.}~\bibnamefont{Matyjasek}} \bibnamefont{and}
  \bibinfo{author}{\bibfnamefont{P.}~\bibnamefont{Sadurski}},
  \bibinfo{journal}{Acta Phys. Polon.} \textbf{\bibinfo{volume}{B45}},
  \bibinfo{pages}{2027} (\bibinfo{year}{2014}).
\bibitem[{\citenamefont{Quinta et~al.}(2018)\citenamefont{Quinta, Flachi, and
  Lemos}}]{Quinta0}
\bibinfo{author}{\bibfnamefont{G.~M.} \bibnamefont{Quinta}},
  \bibinfo{author}{\bibfnamefont{A.}~\bibnamefont{Flachi}}, \bibnamefont{and}
  \bibinfo{author}{\bibfnamefont{J.~P.} \bibnamefont{Lemos}},
  \bibinfo{journal}{Phys. Rev.} \textbf{\bibinfo{volume}{D97}},
  \bibinfo{pages}{025023} (\bibinfo{year}{2018}).
\bibitem[{\citenamefont{Taylor and Breen}(2017)}]{Taylor1}
\bibinfo{author}{\bibfnamefont{P.}~\bibnamefont{Taylor}} \bibnamefont{and}
  \bibinfo{author}{\bibfnamefont{C.}~\bibnamefont{Breen}},
  \bibinfo{journal}{Phys. Rev.} \textbf{\bibinfo{volume}{D96}},
  \bibinfo{pages}{105020} (\bibinfo{year}{2017}).
\bibitem[{\citenamefont{Taylor and Breen}(2016)}]{Taylor2}
\bibinfo{author}{\bibfnamefont{P.}~\bibnamefont{Taylor}} \bibnamefont{and}
  \bibinfo{author}{\bibfnamefont{C.}~\bibnamefont{Breen}},
  \bibinfo{journal}{Phys. Rev.} \textbf{\bibinfo{volume}{D94}},
  \bibinfo{pages}{125024} (\bibinfo{year}{2016}).
\bibitem[{\citenamefont{Flachi et~al.}(2016)\citenamefont{Flachi, Quinta, and
  Lemos}}]{Flachi2}
\bibinfo{author}{\bibfnamefont{A.}~\bibnamefont{Flachi}},
  \bibinfo{author}{\bibfnamefont{G.~M.} \bibnamefont{Quinta}},
  \bibnamefont{and} \bibinfo{author}{\bibfnamefont{J.~P.~S.}
  \bibnamefont{Lemos}}, \bibinfo{journal}{Phys. Rev.}
  \textbf{\bibinfo{volume}{D94}}, \bibinfo{pages}{105001}
  (\bibinfo{year}{2016}).
\bibitem[{\citenamefont{Breen et~al.}(2015)\citenamefont{Breen, Hewitt,
  Ottewill, and Winstanley}}]{Breen}
\bibinfo{author}{\bibfnamefont{C.}~\bibnamefont{Breen}},
  \bibinfo{author}{\bibfnamefont{M.}~\bibnamefont{Hewitt}},
  \bibinfo{author}{\bibfnamefont{A.~C.} \bibnamefont{Ottewill}},
  \bibnamefont{and}
  \bibinfo{author}{\bibfnamefont{E.}~\bibnamefont{Winstanley}},
  \bibinfo{journal}{Phys. Rev.} \textbf{\bibinfo{volume}{D92}},
  \bibinfo{pages}{084039} (\bibinfo{year}{2015}).
\bibitem[{\citenamefont{Page}(1982)}]{Page}
\bibinfo{author}{\bibfnamefont{D.~N.} \bibnamefont{Page}},
  \bibinfo{journal}{Phys. Rev.} \textbf{\bibinfo{volume}{D25}},
  \bibinfo{pages}{1499} (\bibinfo{year}{1982}).
\bibitem[{\citenamefont{Frolov and Zelnikov}(1987)}]{Frolov:1987gw}
\bibinfo{author}{\bibfnamefont{V.~P.} \bibnamefont{Frolov}} \bibnamefont{and}
  \bibinfo{author}{\bibfnamefont{A.~I.} \bibnamefont{Zelnikov}},
  \bibinfo{journal}{Phys. Rev.} \textbf{\bibinfo{volume}{D35}},
  \bibinfo{pages}{3031} (\bibinfo{year}{1987}).
\bibitem[{\citenamefont{Frolov and Zelnikov}(1984)}]{FZ}
\bibinfo{author}{\bibfnamefont{V.~P.} \bibnamefont{Frolov}} \bibnamefont{and}
  \bibinfo{author}{\bibfnamefont{A.~I.} \bibnamefont{Zelnikov}},
  \bibinfo{journal}{Phys. Rev.} \textbf{\bibinfo{volume}{D29}},
  \bibinfo{pages}{1057} (\bibinfo{year}{1984}).
\bibitem[{\citenamefont{Frolov}(1986)}]{Frolovhab}
\bibinfo{author}{\bibfnamefont{V.~P.} \bibnamefont{Frolov}},
  \bibinfo{journal}{Trudy Fiz. Inst. Lebedev.} \textbf{\bibinfo{volume}{169}},
  \bibinfo{pages}{3} (\bibinfo{year}{1986}).
\bibitem[{\citenamefont{Matyjasek}(1996)}]{Matyjasek_masslessIHH}
\bibinfo{author}{\bibfnamefont{J.}~\bibnamefont{Matyjasek}},
  \bibinfo{journal}{Phys. Rev.} \textbf{\bibinfo{volume}{D53}},
  \bibinfo{pages}{794} (\bibinfo{year}{1996}).
\bibitem[{\citenamefont{Matyjasek}(2000)}]{Matyjasek1}
\bibinfo{author}{\bibfnamefont{J.}~\bibnamefont{Matyjasek}},
  \bibinfo{journal}{Phys. Rev.} \textbf{\bibinfo{volume}{D61}},
  \bibinfo{pages}{124019} (\bibinfo{year}{2000}).
\bibitem[{\citenamefont{Matyjasek}(2001)}]{Matyjasek2}
\bibinfo{author}{\bibfnamefont{J.}~\bibnamefont{Matyjasek}},
  \bibinfo{journal}{Phys. Rev.} \textbf{\bibinfo{volume}{D63}},
  \bibinfo{pages}{084004} (\bibinfo{year}{2001}).
\bibitem[{\citenamefont{Taylor et~al.}(2000)\citenamefont{Taylor, Hiscock, and
  Anderson}}]{TPaul}
\bibinfo{author}{\bibfnamefont{B.~E.} \bibnamefont{Taylor}},
  \bibinfo{author}{\bibfnamefont{W.~A.} \bibnamefont{Hiscock}},
  \bibnamefont{and} \bibinfo{author}{\bibfnamefont{P.~R.}
  \bibnamefont{Anderson}}, \bibinfo{journal}{Phys. Rev.}
  \textbf{\bibinfo{volume}{D61}}, \bibinfo{pages}{084021}
  (\bibinfo{year}{2000}).
\bibitem[{\citenamefont{Breen and Ottewill}(2010)}]{BreenO}
\bibinfo{author}{\bibfnamefont{C.}~\bibnamefont{Breen}} \bibnamefont{and}
  \bibinfo{author}{\bibfnamefont{A.~C.} \bibnamefont{Ottewill}},
  \bibinfo{journal}{Phys. Rev.} \textbf{\bibinfo{volume}{D82}},
  \bibinfo{pages}{084019} (\bibinfo{year}{2010}).
\bibitem[{\citenamefont{Vermaseren}(2000)}]{jos}
\bibinfo{author}{\bibfnamefont{J.~A.~M.} \bibnamefont{Vermaseren}}
  (\bibinfo{year}{2000}), \eprint{math-ph/0010025}.

\end{thebibliography}

\appendix*
\section{The general results }
In this appenedix we present in a tabular form our general results for 
the massive quantized field in the spacetime of the static and spherically 
symmetric black holes.
Making use of the formulas~(\ref{og1}) and (\ref{og2}) and the informations 
contained in
the tables
the vacuum polarization $\langle \phi^{2} \rangle_{D}$ can easily be 
reconstructed.
\begin{table}[h!]
 \centering
 \begin{tabular}{|c|c|ccc|ccc|}
  \hline
 & &  \multicolumn{3}{c}{$D=4$} & \multicolumn{3}{c|}{$D=5$} \\ \hline
   $k$ & $F_{k}(r)$ & $\alpha_{k}^{0} $ & $\alpha_{k}^{1}$ & $\alpha_{k}^{2}$& 
$\alpha_{k}^{0}$ & $\alpha_{k}^{1}$ & $\alpha_{k}^{2}$ \\ \hline
 \hline
 1 & $\dfrac{1}{r^4}$ & 1/15 & -2/3 & 2 & 1/2 & -6 & 18 \\
 2 & $\dfrac{f}{r^4}$ & 0 & 2/3 & -4 & -1 & 12 & -36 \\
 3 & $\dfrac{f^2}{r^4}$ & -1/15 & 0 & 2 & 1/2 & -6 & 18 \\
 4 & $\dfrac{f'}{r^3}$ & -1/3 & 10/3 & -8 & -4/3 & 14 & -36 \\
 5 & $\dfrac{f f'}{r^3}$ & 7/15 & -4 & 8 & 26/15 & -16 & 36 \\
 6 & $\dfrac{f'^2}{r^2}$ & 13/45 & -3 & 8 & 59/120 & -6 & 18 \\
 7 & $\dfrac{f''}{r^2}$ & -1/18 & 2/3 & -2 & -1/6 & 2 & -6 \\
 8 & $\dfrac{f f''}{r^2}$ & -1/90 & -1/3 & 2 & -7/30 & 0 & 6 \\
 9 & $\dfrac{f' f''}{r}$ & -1/30 & -2/3 & 4 & -1/20 & -1 & 6 \\
 10 & $f''^2$ & 1/60 & -1/6 & 1/2 & 1/60 & -1/6 & 1/2 \\
 11 & $\dfrac{f f^{(3)}}{r}$ & -1/5 & 1 & 0 & -3/10 & 3/2 & 0 \\
 12 & $f^{(3)} f'$ & -1/30 & 1/6 & 0 & -1/30 & 1/6 & 0 \\
 13 &$ f f^{(4)}$ & -1/30 & 1/6 & 0 & -1/30 & 1/6 & 0 \\
 \hline
\end{tabular}
\caption{The functions $F_{k}(r)$ and the coefficients $\alpha_{k}^{i}$
of the massive scalar field in $(D=4)$ and $(D=5)$-dimensional static,
spherically symmetric black hole.}
\label{ta2a}
 \end{table}
 
 \begin{table}
 \centering
 \begin{tabular}{|c|c|c|c|c|c|}
  \hline
  $k$ & $F_{k}(r)$ & $k$ & $F_{k}(r) $& $k$ & $F_{k}(r)$ \\ \hline
$ 1 $ & $  \dfrac{1}{r^6} $ & $ 13 $ & $  \dfrac{f^2 f''}{r^4} $ & $ 25 $ & $  
\dfrac{f^{(3)} f'^2}{r} $ \\
$ 2 $ & $  \dfrac{f}{r^6} $ & $ 14 $ & $  \dfrac{f' f''}{r^3} $ & $ 26 $ & $  
\dfrac{f f^{(3)} f''}{r} $ \\
$ 3 $ & $  \dfrac{f^2}{r^6} $ & $ 15 $ & $  \dfrac{f f' f''}{r^3} $ & $ 27 $ & $ 
f^{(3)} f' f'' $ \\
$ 4 $ & $  \dfrac{f^3}{r^6} $ & $ 16 $ & $  \dfrac{{f'}^2 f''}{r^2} $ & $ 28 $ & 
$ f {f^{(3)}}^{2} $ \\
$ 5 $ & $  \dfrac{f'}{r^5} $ & $ 17 $ & $  \dfrac{f''^2}{r^2} $ & $ 29 $ & $  
\dfrac{f f^{(4)}}{r^2} $ \\
$ 6 $ & $  \dfrac{f f'}{r^5} $ & $ 18 $ & $  \dfrac{f f''^2}{r^2} $ & $ 30 $ & $ 
 \dfrac{f^2 f^{(4)}}{r^2} $ \\
$ 7 $ & $  \dfrac{f^2 f'}{r^5} $ & $ 19 $ & $  \dfrac{f' f''^2}{r} $ & $ 31 $ & 
$  \dfrac{f f^{(4)} f'}{r} $ \\
$ 8 $ & $  \dfrac{f'^2}{r^4} $ & $ 20 $ & $ f''^3 $ & $ 32 $ & $ f^{(4)} f'^2 $ 
\\
$ 9 $ & $  \dfrac{f f'^2}{r^4} $ & $ 21 $ & $  \dfrac{f f^{(3)}}{r^3} $ & $ 33 $ 
& $ f f^{(4)} f'' $ \\
$ 10 $ & $  \dfrac{f'^3}{r^3} $ & $ 22 $ & $  \dfrac{f^2 f^{(3)}}{r^3} $ & $ 34 
$ & $  \dfrac{f^2 f^{(5)}}{r} $ \\
$ 11 $ & $  \dfrac{f''}{r^4} $ & $ 23 $ & $  \dfrac{f^{(3)} f'}{r^2} $ & $ 35 $ 
& $ f f^{(5)} f' $ \\
$ 12 $ & $  \dfrac{f f''}{r^4} $ & $ 24 $ & $  \dfrac{f f^{(3)} f'}{r^2} $ & $ 
36 $ & $ f^2 f^{(6)} $ \\
 \hline
\end{tabular}
\caption{The functions $F_{k}(r)$ of the quantized massive field in the 
spacetime
of the static and spherically-symmetric black holes in $D=6$ and $7.$}
\label{ta2}
 \end{table}

\begin{table}
 \centering	
 \begin{tabular}{|c| cccc | c | cccc|}
  \hline
 $k$ & $\alpha_{k}^{0}$ & $\alpha_{k}^{1}$ & $\alpha_{k}^{2}$ & $\alpha_{k}^{3}$ 
& $k$ &
 $\alpha_{k}^{0}$ & $\alpha_{k}^{1}$ & $\alpha_{k}^{2}$ & $\alpha_{k}^{3}$ \\
 \hline 
 1 & $74/63$ & -116/5 & 144 & -288 & 19 & 2/315 & -1/15 & -2/3 & 4 \\
 2 & -58/15 & 356/5 & -432 & 864 & 20 & -1/630 & 1/60 & -1/12 & 1/6 \\
 3 & 58/15 & -356/5 & 432 & -864 & 21 & -11/15 & 116/15 & -20 & 0 \\
 4 & -74/63 & 116/5 & -144 & 288 & 22 & 92/105 & -42/5 & 20 & 0 \\
 5 & -58/15 & 952/15 & -336 & 576 & 23 & -1/15 & 11/15 & -2 & 0 \\
 6 & 164/15 & -2384/15 & 752 & -1152 & 24 & 149/630 & -229/45 & 56/3 & 0 \\
 7 & -778/105 & 1456/15 & -416 & 576 & 25 & -8/315 & -8/45 & 4/3 & 0 \\
 8 & 116/45 & -204/5 & 216 & -384 & 26 & -4/315 & -29/45 & 10/3 & 0 \\
 9 & -1103/315 & 2224/45 & -236 & 384 & 27 & 1/210 & -1/20 & 1/6 & 0 \\
 10 & -2/315 & 184/45 & -112/3 & 256/3 & 28 & 1/840 & -1/60 & 1/12 & 0 \\
 11 & -29/90 & 89/15 & -36 & 72 & 29 & -1/15 & 11/15 & -2 & 0 \\
 12 & -1/45 & -68/15 & 52 & -144 & 30 & -5/42 & 2/15 & 2 & 0 \\
 13 & 65/126 & -11/5 & -16 & 72 & 31 & -44/315 & 16/45 & 4/3 & 0 \\
 14 & -1/15 & -44/15 & 36 & -96 & 32 & -1/140 & 1/30 & 0 & 0 \\
 15 & 683/315 & -232/15 & 4 & 96 & 33 & -1/420 & -1/60 & 1/6 & 0 \\
 16 & 131/630 & -64/45 & -14/3 & 32 & 34 & -2/35 & 4/15 & 0 & 0 \\
 17 & 1/30 & -8/15 & 3 & -6 & 35 & -1/70 & 1/15 & 0 & 0 \\
 18 & 89/630 & -109/45 & 7 & 6 & 36 & -1/280 & 1/60 & 0 & 0 \\
 \hline
\end{tabular}
\caption{The coefficients $\alpha_{k}$ of the quantized massive field in the 
spacetime
of the static and spherically-symmetric black holes in $D=6.$}
\label{ta3}
 \end{table}

 \begin{table}
 \centering
 \begin{tabular}{|c| cccc | c | cccc|}
  \hline
 $k$ & $\alpha_{k}^{0}$ & $\alpha_{k}^{1}$ & $\alpha_{k}^{2}$ & $\alpha_{k}^{3}$ 
& $k$ &
 $\alpha_{k}^{0}$ & $\alpha_{k}^{1}$ & $\alpha_{k}^{2}$ & $\alpha_{k}^{3}$ \\
 \hline
  1 & 16/3 & -320/3 & 2000/3 & -4000/3 & 19 & 1/126 & -1/12 & -5/6 & 5 \\
 2 & -64/3 & 1120/3 & -6400/3 & 4000 & 20 & -1/630 & 1/60 & -1/12 & 1/6 \\
 3 & 80/3 & -1280/3 & 6800/3 & -4000 & 21 & -14/9 & 149/9 & -130/3 & 0 \\
 4 & -32/3 & 160 & -800 & 4000/3 & 22 & 793/504 & -599/36 & 130/3 & 0 \\
 5 & -112/9 & 1880/9 & -3400/3 & 2000 & 23 & -1/9 & 11/9 & -10/3 & 0 \\
 6 & 104/3 & -4622/9 & 2500 & -4000 & 24 & 61/252 & -68/9 & 30 & 0 \\
 7 & -5843/252 & 1855/6 & -4100/3 & 2000 & 25 & -2/63 & -2/9 & 5/3 & 0 \\
 8 & 199/36 & -1697/18 & 1600/3 & -1000 & 26 & -1/63 & -29/36 & 25/6 & 0 \\
 9 & -2659/504 & 569/6 & -1625/3 & 1000 & 27 & 1/210 & -1/20 & 1/6 & 0 \\
 10 & 19/126 & 71/12 & -200/3 & 500/3 & 28 & 1/840 & -1/60 & 1/12 & 0 \\
 11 & -8/9 & 148/9 & -100 & 200 & 29 & -1/9 & 11/9 & -10/3 & 0 \\
 12 & -4/3 & 2/9 & 340/3 & -400 & 30 & -115/504 & 13/36 & 10/3 & 0 \\
 13 & 803/252 & -127/6 & -40/3 & 200 & 31 & -11/63 & 4/9 & 5/3 & 0 \\
 14 & -1/6 & -17/3 & 220/3 & -200 & 32 & -1/140 & 1/30 & 0 & 0 \\
 15 & 1093/252 & -209/6 & 25 & 200 & 33 & -1/420 & -1/60 & 1/6 & 0 \\
 16 & 97/336 & -53/24 & -20/3 & 50 & 34 & -1/14 & 1/3 & 0 & 0 \\
 17 & 1/18 & -8/9 & 5 & -10 & 35 & -1/70 & 1/15 & 0 & 0 \\
 18 & 95/504 & -23/6 & 35/3 & 10 & 36 & -1/280 & 1/60 & 0 & 0 \\
 \hline
\end{tabular}
\caption{The coefficients $\alpha_{k}$ of the quantized massive field in the 
spacetime
of the static and spherically-symmetric black holes in $D=7.$}
\label{ta4}
 \end{table}

\end{document}